\documentstyle[12pt,fleqn]{article}
\input psfig
\begin{document}
\begin{titlepage}
\null\vspace{-62pt}
\begin{flushright}
{\footnotesize
FERMILAB--Pub--94/053-A\\
astro-ph/9403004\\
February 1994 \\
Submitted to {\em Phys.\ Rev.\ Lett}}
\end{flushright}
\renewcommand{\thefootnote}{\fnsymbol{footnote}}
\vspace{0.25in}
\def\pag{{{\langle P \rangle}}}
\begin{center}
{\Large \bf Testing inflation with the \\ cosmic microwave background}\\
\vspace{1.0cm}
Scott Dodelson\footnote{Electronic mail: {\tt dodelson@virgo.fnal.gov}}\\
{\em NASA/Fermilab Astrophysics Center\\
Fermi National Accelerator Laboratory, Batavia, IL~~60510}\\
\vspace{0.4cm}
Lloyd Knox\footnote{Electronic mail: {\tt knox@oddjob.uchicago.edu}}\\
{\em Department of Physics, Enrico Fermi Institute\\
The University of Chicago, Chicago, IL~~ 60637}\\
\vspace{0.4cm}
Edward W.\ Kolb\footnote{Electronic mail: {\tt rocky@fnas01.fnal.gov}}\\
{\em NASA/Fermilab Astrophysics Center\\
Fermi National Accelerator Laboratory, Batavia, IL~~60510, and\\
Department of Astronomy and Astrophysics, Enrico Fermi Institute\\
The University of Chicago, Chicago, IL~~ 60637}\\

\end{center}
\baselineskip=24pt

\vspace{.5cm}

\begin{quote}
\hspace*{2em}  Cosmic microwave background (CMB) anisotropy may result from
both scalar and tensor perturbations. For a sufficiently narrow range of
angular scales, CMB perturbations can be characterized by four
parameters. Results from the Cosmic Background Explorer fix one
combination of the parameters, reducing the parameters to three.  If
CMB perturbations are from inflation, there is an additional relation,
reducing the parameters to two.  An appropriate combination of a
medium-angle and a small-angle CMB observation can test the inflation
hypothesis because inflation cannot explain a high signal in one
experiment and a low signal in the other.
\vspace*{12pt}

PACS number(s): 98.80.Cq, 98.70.Vc, 98.80.Es

\renewcommand{\thefootnote}{\arabic{footnote}}
\addtocounter{footnote}{-2}
\end{quote}
\end{titlepage}
\addtocounter{page}{1}
\newpage
\baselineskip=24pt

\def\re#1{{[\ref{#1}]}}
\def\be{\begin{equation}}
\def\ee{\end{equation}}
\def\bea{\begin{eqnarray}}
\def\eea{\end{eqnarray}}
\def\ns{{n_S}}
\def\nt{n_T}
\def\sone{\delta T^{(1)}}
\def\stwo{\delta T^{(2)}}
\def\wone{W_l^{(1)}}
\def\wtwo{W_l^{(2)}}
\def\pone{P_1}
\def\ptwo{P_2}

Guth pointed out that a rapid expansion early in the history of our
Universe could solve several problems of the standard cosmology, the
most notable being the horizon and flatness problems \re{guth}.  It
was realized shortly after the initial proposal that inflation does
more than provide a smooth Universe: It also generates small
perturbations on the smooth background \re{infper}. These small
perturbations may well be the ones that grow via gravitational
instability into the diversity of structures we see today in our
Universe.

Most successful models of inflation involve the dynamics of a weakly
coupled scalar field that slowly evolves (rolls) under the influence
of some scalar potential.  This scalar field is called the {\em
inflaton}.  While the inflaton evolves during inflation, perturbations
in the energy density arise as the result of quantum mechanical
fluctuations of the inflaton field in de Sitter space.  These
perturbations in the energy density correspond to {\em scalar} metric
fluctuations.  In addition to the scalar metric fluctuations there
should be {\em tensor} perturbations caused by de Sitter space
fluctuations of the metric tensor \re{tensor}.

In principle, one must specify the amplitude of the tensor and scalar
perturbations on all length scales, each of which enters the horizon
at different cosmic times.  However, if the inflaton evolves slowly
during inflation (as it must in order that its potential energy
dominate the energy density) then the resulting scalar perturbations
should approximately have the Harrison--Zel'dovich spectrum; the
Fourier transform of their variance is a simple power law with
exponent $\ns=1$.  In the same slow-roll limit the Fourier transform
of the variance of the tensor perturbations is also a power law, with
exponent $\nt=0$.

Of course the fact that the scalar field must evolve during inflation
means that the resulting spectra won't have exactly the above simple
form.  However since the field must evolve slowly, over a limited
range of length scales (such as the length scales probed by CMB
experiments) it should be possible to describe the scalar and tensor
perturbations as power laws, with exponents not too different from the
Harrison--Zel'dovich values. Therefore we will assume that the spectra
can be described by four parameters: the amplitudes and spectral
indices of the scalar and tensor spectra.

The purpose of this Letter is to show that inflation is testable even
though there are four parameters which describe the scalar and tensor
spectra. In particular we argue that a good test of inflation can be
constructed by combining information from three sets of cosmic
microwave background anisotropy experiments: COBE \re{COBE} at large
angular scales; one at medium angular scales, and a third at small
angular scales.  At first glance this seems a hollow claim, for there
are four free parameters in the scalar and tensor spectra, and with
four free parameters one should easily be able to fit three
experimental results. How can just three experiments test inflation?

Fortunately, inflation does make a generic prediction; namely a
relationship between the shape of the tensor spectrum and its
amplitude. Thus, there are only three free parameters if the
perturbations arise from inflation.  Suppose we fix one of these
parameters with the COBE result. As we vary the other two, we can
indeed significantly change the expected signal in both the small and
the medium angle experiments.  However, we find that {\it the signal
in the small-angle experiment scales in a predictable way as a
function of the signal in the medium-angle experiment.} So once the
signal in one of these is fixed, the signal in the other is
unambiguously determined.  We can think of this abstractly as a
mapping of the two-dimensional space of the two remaining free
parameters after COBE normalization onto the two-dimensional space of
signals in the medium- and small-angle experiments.  If the mapping
were onto, so that any point in the signal space could be reached by a
point in the parameter space, then inflation could never be disproved
by this set of experiments.  Our claim is that the two dimensional
parameter space is mapped into a single one-dimensional line in signal
space.  A combination of experimental results not falling on this line
cannot result from inflation, so there is a whole range of
experimental results which can disprove inflation.

Before discussing more precisely what we mean by ``medium-'' and
``small-'' angle experiments, we must introduce some notation.  It has
become standard to decompose the angular correlation function into a
sum of Legendre polynomials:
\be \label{ctheta}
C(\theta)  \equiv \langle \delta T(\theta) \delta T(0) \rangle
	= \sum_{l=2}^\infty {2l+1\over 4\pi} C_l P_l(\cos\theta).
\ee
Recall that in this expansion, small $l$ corresponds to large angles
while large $l$ corresponds to small angles.  Experiments do not
directly measure $C(\theta)$, but rather measure some convolution of
the $C_l$'s with a window function \re{window}, $W_l$, determined by
the particular experiment.  Thus, the predicted variance in a given
experiment is defined as
\be
\label{dtobs}
\langle \delta T_{\rm exp}^2 \rangle = \sum_{l=2}^\infty
{2l+1\over 4\pi} C_l W_l.
\ee

To determine the predictions of a given theory, one must calculate the
set of $C_l$'s it predicts \re{clcal}.  Let's briefly review the
steps involved in such a calculation: (i) perturb the Einstein and
Boltzmann equations about the standard zero-order solutions (the
Robertson-Walker metric with homogeneous and isotropic distributions
of photons, neutrinos, ordinary matter, and dark matter); (ii) Fourier
transform these equations to express the perturbations $\Delta$ in
terms of wavenumber $k$, time $t$, and in the case of photons and
neutrinos, the angle $\theta$ between the wavenumber and momentum;
(iii) Expand the perturbations to the photons and neutrinos in terms
of Legendre polynomials so that the angular dependence,
$\Delta(\theta)$, is replaced by the coefficients, $\Delta_l$; (iv)
Evolve these perturbed quantities starting from initial conditions
deep in the radiation era: $(\delta\rho/\rho)_S (k,t_{\rm init})
\propto k^{\ns/2} ; (\delta\rho/\rho)_T (k,t_{\rm init}) \propto
k^{(\nt-3)/2}$ where $\ns=1; \nt=0$ for the Harrison--Zel'dovich
spectra; (v) Determine the $C_l$'s due to both scalar and tensor modes
today by integrating $C_{l,(S,T)} \propto
\int d^3k \vert \Delta_{l,(S,T)} (t_0)\vert^2$;
(vi) Add the two contributions \re{gold}:
$C_l = C_{l,S} + C_{l,T}$. The proportional signs
in steps (iv) and (v) are an indication that we do
not know the normalization of either mode.  We can fix
one such parameter, say $C_{2,S}$, by using the COBE result
\re{adams}. The three remaining unknowns are
$R\equiv C_{2,T}/C_{2,S}$; $\ns$, and $\nt$.

Now that we have defined the relevant parameters we can discuss the
prediction of inflation.  The tensor-to-scalar energy density ratio
\re{clearup}, $r$, as well as the spectral indices can be expressed
in terms of the derivatives of the expansion rate, $H$, during
inflation.  In the limit that $H$ is constant during inflation,
$\left[ r,\ \nt,\ \ns \right] = [0,0,1]$.  However in slow-roll
inflation $H$ changes in time.  Using the value of the inflaton field
as the time variable [$\phi=\phi(t)$], then one can calculate $\left[
r,\ \nt,\ \ns \right]$ as a function of $\epsilon\equiv
2[H'(\phi)/H(\phi)]/\kappa^2$, $\eta \equiv
2[H''(\phi)/H'(\phi)]/\kappa^2$, and $\xi\equiv
2[H'''(\phi)/H'(\phi)]/\kappa^2$. Here $\kappa^2 \equiv 8\pi/m_{\rm
Planck}^2$ and prime denotes $d/d\phi$.  If ($\epsilon,\ \eta,\ \xi)$
are less than unity, then to first order in these parameters
\re{slowroll}
\be
\left[ r,\ \nt,\ 1-\ns \right] = \left[ \frac{25}{4}\,
2\epsilon,\ -2\epsilon,\ 2\epsilon\left(2-\frac{\eta}{\epsilon}
\right) \right]
\ee
Therefore, to first order $r=-6.25 \nt$.  A scalar-to-tensor ratio of
$r=-6.25\nt$ results in $R\equiv C_{2,T}/C_{2,S}=-7\nt$.

The set of $C_l$'s predicted by inflation is therefore dependent on
two free parameters: $C_l = C_l(\ns,R=-7\nt)$. Fig.\ 1 shows the
$C_l$'s for several different values of these two parameters.  The
solid curve is for standard inflation [$(R,\ \ns) = (0,\ 1)$].  As
$\ns$ increases the signal increases, the effect being greatest on the
smallest scales.  So the dashed curve, which is for $(R,\ \ns) = (0,\
1.25)$, is higher than the standard one.  As $R$ increases, the signal
at small angular scales [large $l$] goes down: $(R,\ \ns) = (2,1)$
produces the dotted curve in Fig.\ 1. The point is that the COBE
signal [which comes from $l<20$] is partly due to tensor modes in this
case, thereby reducing the amplitude of the scalar component.  The
tensor contribution drops off after $l=100$ (physically this is
because gravity waves redshift once they enter the horizon; $l=100$ is
roughly the scale of the horizon at decoupling).  So once $l>150$, all
that is left is the reduced scalar contribution.  Note however that
the signal in the medium angle range $l\sim 50$ also decreases. This
is a consequence of the inflationary prediction $\nt = -R/7$. Since
$R=2$ in this case, the tensor spectrum is tilted so as to fall off
rapidly with increasing $l$. We can thus imagine that while
alternative models of inflation may change the signal in either a
small or a medium angle experiment, it will be very hard to change the
signals in opposite directions.  This observation is the basis for the
test we will shortly describe.

Before getting into the details of the test, we should spell out
our assumptions. The $C_l$'s in Fig.\ 1 were generated assuming
cold dark matter; zero cosmological constant; standard ionization
history; Hubble constant today $H_0 = 50$ km sec$^{-1}$
Mpc$^{-1}$; and the fraction of critical density in baryons
today, $\Omega_B = 0.05$. Varying any of these parameters
can lead to significant changes in the $C_l$'s \re{stein}.
Our philosophy is that these parameters, while not particularly
well-determined today, are very likely to be determined in
the future {\it by experiments other than microwave anisotropy
experiments} \re{ion}. Therefore, we feel it is unlikely that
our lack of knowledge about these parameters will be the stumbling
block keeping us from testing inflation.

Now for a method to test inflation. Imagine two anisotropy
experiments, one with a medium-angle filter, $\wone = 1$ for $30<l<90$
and zero otherwise; the other with a small-angle filter, $\wtwo = 1$
for $130<l<300$ and zero otherwise. Then the predicted signal in each
experiment is
\bea
\label{signal}
\sone(\ns,\nt,R) &=& \left[
\sum_{l=30}^{90} {2l+1\over 4\pi} C_l(\ns,\nt,R) \right]^{1/2}
\nonumber \\
\stwo(\ns,\nt,R) &=& \left[
\sum_{l=130}^{300} {2l+1\over 4\pi} C_l(\ns,\nt,R) \right]^{1/2}.
\eea
The first experiment would sample both scalar and tensor modes, while
the second would sample only scalar modes. We have explicitly
indicated that the signal in these experiments depends on the values
of $(\ns,\nt,R)$.  The set of parameters allowed by inflation can now
be mapped onto these two signals. Fig.\ 2 shows such a mapping for
$-0.5 < \ns < 1.5;~ R < 3.5 ; \nt = -R/7$. A larger range would not be
consistent with our use of equation 3, which is true only to first
order in the slow-roll parameters, $\epsilon$ and $\eta$. The
important point is that this whole region of parameter space allowed
by inflation is mapped onto a very narrow region, almost a line, in
signal space \re{natural}. There is some scatter off this line,
particularly if the signals are low. However, this low signal regime
comes from $\ns < 0.5$; such small values of $\ns$, as we discuss
later, are highly unlikely given other data.

Inflation therefore predicts small deviations from this ``line'' in
signal space.  We can quantify this further by defining a linear
combination of the signals such that one of the new variables is the
distance from the locus of inflation predictions. A good measure of
this distance is
\be
\label{defd}
D  \equiv  \alpha \left[
	{\sone \over \sone(\ns=1;R=0)} - 1 \right]
	- \beta \left[
	{\stwo \over \stwo(\ns=1;R=0)} - 1 \right]
\ee
where
\begin{eqnarray*} \label{pardef}
\alpha &\equiv& {1\over \sqrt{1+\gamma^2}}
\quad ; \qquad
\beta \equiv {1\over \sqrt{1+\gamma^{-2}}}
\\
\gamma &\equiv& { 1 - \sone(\ns=.5,R=0)/\sone(\ns=1,R=0)
		\over 1 - \stwo(\ns=.5,R=0)/\sone(\ns=1,R=0) } .
\end{eqnarray*}
Note that $\alpha$, $\beta$, and $\gamma$ can be calculated for any
pair of filter functions. For the simple square ones
\re{ofil} we have chosen, $\alpha = 0.77$ and $\beta = 0.63$.
So $D$ is roughly the difference between the signals in the
small and medium angle experiments, each of which
is normalized by its standard value at $\ns=1$. The fact that
inflation predicts $D\simeq 0$ means that {\it inflation cannot
explain a high signal in one of these experiments and a low one in the other}.

Fig.\ 3 shows $D$ as a function of $\ns$ and $R=-7\nt$.  As mentioned
above, $D$ does begin to deviate from zero, but only in
``non-physical'' regions in the parameter space. A way to quantify
this is to note that $\sigma_8$, the {\it rms} mass fluctuation in
spheres of radius $8h^{-1}{\rm Mpc}$, must certainly be greater
\re{sigmaeight} than about $1/3$, while the light region in Fig.\ 3
has $\sigma_8<1/3$.  [On the color version, colors redder than yellow
in Fig.\ 3 have $\sigma_8<1/3$.]  So the value of $D$ in these regions
is irrelevant. In the ``allowed'' range, $D$ is always less than
$0.06$. When $\ns > 1$, $D$ can become negative, as small as $-0.1$;
however, such large values of $\ns$ are also thought to be unlikely
because they would produce too much structure on scales of about $1$
Mpc.

We have defined a new parameter $D$ [Eq.\ (\ref{defd})] which combines
information from a small and a medium angle experiment.  Many
experiments currently on-line are probing the angular regimes
necessary to evaluate $D$.  Viable models of inflation predict
$D<0.06$. It remains to be seen what values of $D$ are predicted by
other cosmological theories, such as those with non-Gaussian seeded
perturbations.

It is a pleasure to thank Michael Turner for helpful programming
suggestions.  This work was supported in part by the DOE and NASA
grant NAGW--2381 at Fermilab.

\frenchspacing
\begin{picture}(400,50)(0,0)
\put (50,0){\line(350,0){300}}
\end{picture}
\vspace{0.25in}

\def\labelenumi{[\theenumi]}

\begin{enumerate}

\item\label{guth} A. H. Guth, Phys. Rev. {\bf D23}, 347 (1981).

\item\label{infper} A. H. Guth and S.-Y. Pi, Phys.
Rev. Lett. {\bf 49}, 1110 (1982); S. W. Hawking, Phys.
 Lett. {\bf 115B}, 295 (1982);  A. A. Starobinskii, Phys.
Lett. {\bf 117B}, 175 (1982); J. M. Bardeen, P. J. Steinhardt,
and M. S. Turner, Phys. Rev. {\bf D28}, 697 (1983).

\item\label{tensor} V. A. Rubakov, M. Sazhin, and A.
Veryaskin, Phys. Lett. {\bf 115B}, 189 (1982); R.
Fabbri and M. Pollock, Phys. Lett. {\bf 125B}, 445 (1983);
L. Abbott and M. Wise, Nucl. Phys. {\bf B 244}, 541 (1984);
B. Allen, Phys. Rev. {\bf D37}, 2078 (1988); L. P.  Grishchuk,
Phys. Rev. Lett. {\bf 70}, 2371 (1993).

\item\label{COBE} Smoot, G. F. {\it et al.} 1992, {\it Ap. J.},
{\bf 396}, L1.

\item\label{window} J. R. Bond, G. Efstathiou, P. M. Lubin,
and P. R. Meinhold, Phys. Rev. Lett. {\bf 66}, 2179 (1991); S.
Dodelson and J. M. Jubas, Phys. Rev.  Lett. {\bf 70}, 2224 (1993); M.
White, D. Scott, and J. Silk, to appear in Annu. Rev. Astron.
Astrophys. (1994); M. White and M. Srednicki, astro-ph/9402037 (1994).

\item\label{clcal} The most famous contribution to the
$C_l$'s is the Sachs-Wolfe effect discussed in R. K.  Sachs and A. M.
Wolfe, Astrophs. J.  {\bf 147}, 73 (1967) which accounts for large
scale effects due to perturbations to the metric. The complete
calculation valid even on smaller scales was carried out for theories
with only baryonic matter by P. J. E. Peebles and J. T. Yu, Astrophys.
J.  {\bf 162}, 815 (1970). The calculation was extended to theories
with non-baryonic matter and refined by M. L.  Wilson and J. Silk,
Astrophys. J.  {\bf 243}, 14 (1981); J. R. Bond and G. Efstathiou,
Astrophys. J.  {\bf 285}, L45 (1984); J. R. Bond and G. Efstathiou,
MNRAS {\bf 226}, 655 (1987). For a clear review, see G.  Efstathiou in
{\it Physics of the Early Universe}, edited by J. A. Peacock, A. F.
Heavens, and A. T.  Davies (Edinburgh University Press, Edinburgh,
1990).  All this was for theories with primordial scalar
perturbations.  Theories with tensor perturbations have been recently
analyzed in a similar fashion by R. Crittenden, J. R. Bond, R. L.
Davis, G.  Efstathiou, and P. J. Steinhardt, Phys. Rev. Lett. {\bf
71}, 324 (1993).

\item\label{gold} The possibility that the COBE signal
could be due in part to tensor modes was pointed out by L. Krauss and
M.  White, Phys. Rev. Lett.  {\bf 69}, 869 (1992); R. Davis, {\it
etal.}, {\it ibid.} {\bf 69}, 1856 (1992); D. Salopek, {\it ibid.}
{\bf 69}, 3602 (1992); F. Lucchin, S. Mattarese, and S. Mollerach,
Astrophys. J.  {\bf 401}, L49 (1992); A. Liddle and D. Lyth, Phys.
Lett.  {\bf 291B}, 391 (1992); T. Souradeep and V. Sahni, Mod.  Phys.
Lett. {\bf A7}, 3541 (1992).

\item\label{adams} We follow the normalization procedure
of F. C. Adams,  J. R. Bond, K. Freese, J. A. Frieman,
and A. V. Olinto, Phys. Rev. D {\bf 47}, 426    (1993).
This procedure has recently been refined by E. Wright {\it et al.},
Astrophys. J. {\bf 420}, 1 (1994).  However,
we expect small
errors in the relative normalization of different sets of $C_l$'s.

\item\label{clearup} Note that $r$ is the ratio of energy density
in tensor as compared with scalar modes; this differs slightly
from $R$, the ratio of the ``quadrupoles:'' $C_{2,T}/C_{2,S}$.

\item\label{slowroll} E. D. Stewart and D. H. Lyth, Phys. Lett.
{\bf B302}, 171 (1993); E. J. Copeland, E. W. Kolb, A. R. Liddle, and
J. E. Lidsey, Phys. Rev. Lett. {\bf 71}, 219 (1993); Phys. Rev. D {\bf
48}, 2529 (1993); {\bf 49}, 1840 (1994); M. S. Turner, Phys. Rev. D
{\bf 48}, 5539 (1993); A. R. Liddle and M. S. Turner, ``Second-order
reconstruction of the inflationary potential,'' FNAL-Pub-93/399-A
(unpublished); E. W. Kolb and S. L. Vadas, ``Relating spectral indices
to tensor and scalar ratios in inflation,'' FNAL-Pub-94/046-A,
astro-ph/9403001 (unpublished).

\item\label{stein} The issue of how anisotropies vary with
the the cosmological parameters has been often discussed.
For a recent treatment see J. R. Bond, R. Crittenden, R. L.
Davis, G. Efstathiou, and P. J. Steinhardt, astro-ph/9309041
``Measuring Cosmological Parameters with Cosmic Microwave
Background Experiments.''

\item\label{ion} The ionization history is the one area where
it is not clear what experiments can be done to test the standard
assumption.  Particularly thorny is the issue of late reionization.

\item\label{natural} The low scatter in Figure 2 also means that a
simple model of inflation with no tensor modes [see e.g. Ref.
\re{adams}] is an effective way of probing all inflationary models.
That is, by varying $n$ alone, we can get any set of signals that
might be generated by including tensor modes as well.

\item\label{ofil} We have also tried Gaussian filter functions
centered at $l=60$ and $l=200$. As long as neither extends
much over the $l=100$ boundary,  $\alpha$ and $\beta$ have
the same numerical values given in the text.

\item\label{sigmaeight} In Fig.\ 3, $\sigma_8$ has been
determined assuming cold dark matter; any other form of dark matter
leads to even smaller values of $\sigma_8$.

\end{enumerate}

\newpage

\centerline{\bf FIGURE CAPTIONS}

\vspace{24pt}

\hspace*{0.1in} FIG.\ 1: $C_l$'s predicted for three
different inflation models.  $C_l$'s are in units of $(\mu$K$)^2$.
\vspace{24pt}

\hspace*{0.1in} Fig.\ 2: A mapping from the plane of inflation
predictions into the signal plane of the medium $(\delta T^{(1)})$
 and small $(\delta T^{(2)})$ angle experiment plane. Also shown are
contours of constant $D$, defined in Eq.\ 5, and the error bars [centered
around the $(n=1,R=0)$ prediction] for experiments with $50$ pixels
and unity signal/noise ratio.
\vspace{24pt}

\hspace*{0.1in} Fig.\ 3: The parameter $D$ of Eq.\
(\ref{defd}) as a function of $\ns$ and $R$.
\vspace{24pt}

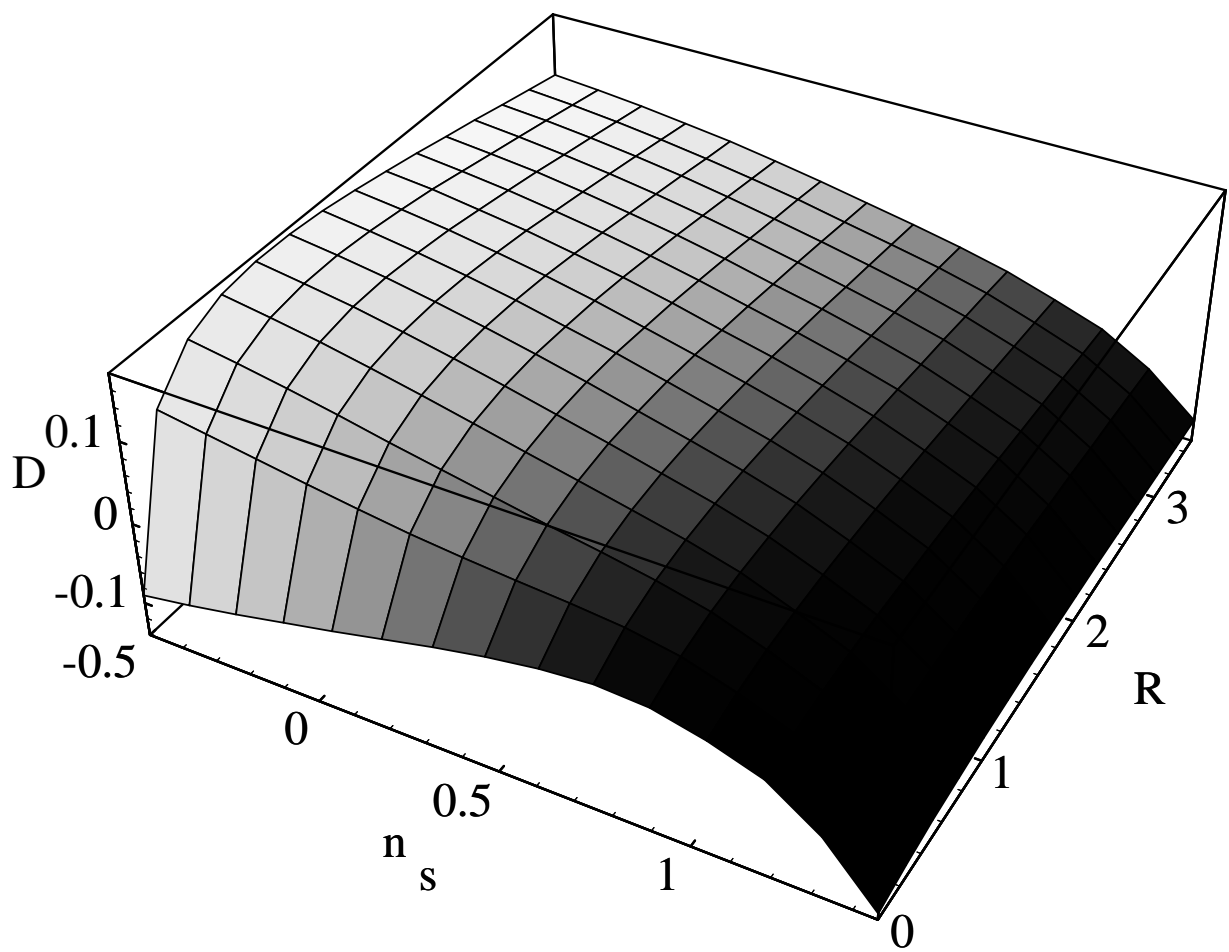
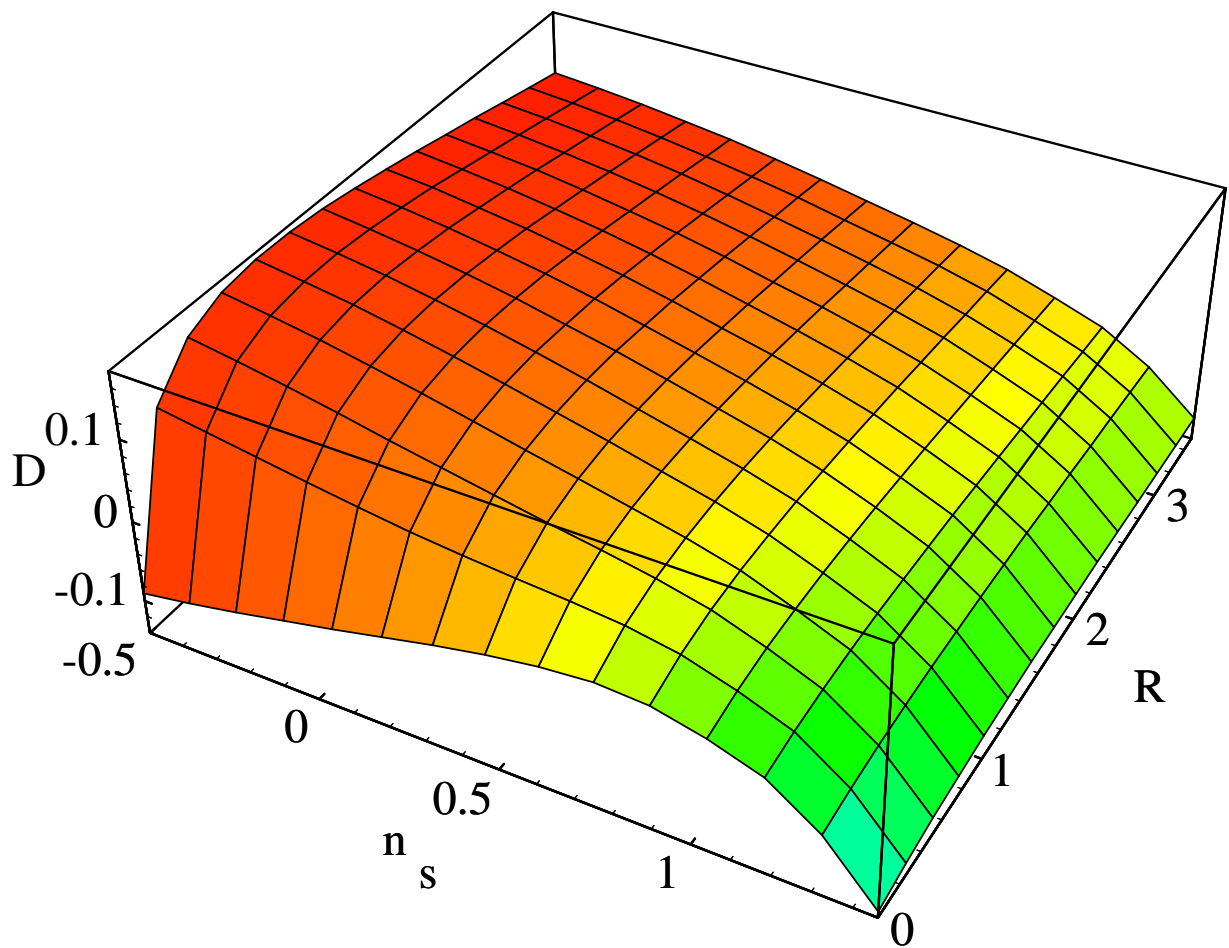
\end{document}